\documentclass[preprint,aps,showkeys]{revtex4}

\usepackage{graphicx}
\usepackage{dcolumn}
\usepackage{bm}

\begin{document}


\title{Tuning the Correlation Decay in the Resistance
\\Fluctuations of Multi-Species Networks}

\author{C. Pennetta}
\affiliation{Dipartimento di Ingegneria dell'Innovazione, 
Universit\`a del Salento and CNISM, Via Arnesano, I-73100, 
Lecce, Italy.} 
\email{cecilia.pennetta@unile.it}

\author{E. Alfinito}
\affiliation{Dipartimento di Ingegneria dell'Innovazione, 
Universit\`a del Salento and CNISM, Via Arnesano, I-73100, 
Lecce, Italy.}

\author{L. Reggiani}
\affiliation{Dipartimento di Ingegneria dell'Innovazione, 
Universit\`a del Salento and CNISM, Via Arnesano, I-73100, 
Lecce, Italy.}

\thanks{Corresponding author e-mail: cecilia.pennetta@unile.it }

\date{\today}

\begin{abstract}
A new network model is proposed to describe the $1/f^\alpha$ resistance noise 
in disordered materials for a wide range of $\alpha$ values ($0< \alpha < 2$).
More precisely, we have considered the resistance fluctuations of a thin
resistor with granular structure in different stationary states: from nearly 
equilibrium up to far from equilibrium conditions. This system has been 
modelled as a network made by different species of resistors, distinguished by 
their resistances, temperature coefficients and by the energies associated
with thermally activated processes of breaking and recovery. 
The correlation behavior of the resistance fluctuations is analyzed as a 
function of the temperature and applied current, in both the frequency and
time domains. For the noise frequency exponent, the model provides 
$0< \alpha <  1$ at low currents, in the Ohmic regime, with $\alpha$ 
decreasing inversely with the temperature, and $1< \alpha <2$ at high 
currents, in the non-Ohmic regime. Since the threshold current associated 
with the onset of nonlinearity also depends on the temperature, the proposed 
model qualitatively accounts for the complicate behavior of $\alpha$ versus 
temperature and current observed in many experiments. Correspondingly, in 
the time domain, the auto-correlation function of the resistance fluctuations 
displays a variety of behaviors which are tuned by the external conditions. 
\end{abstract} 

\maketitle

%
%
 
\section{Introduction} 
The analysis of resistance fluctuations has proved 
to be a very powerful tool for probing various condensed matter systems 
\cite{review,kogan,upon05}, including nanostructures 
\cite{bosman,raychau_wire,soliveres,tarasov,tersoff,appenzeller,leturcq}
and disordered materials, like conductor-insulator composites
\cite{torquato,sahimi,bardhan}, granular systems \cite{torquato,sahimi,grier}, 
porous \cite{torquato,bloom} or amorphous materials 
\cite{kakalios,lust94,alers,weissman_1f}, organic conducting blends 
\cite{planes_1f,carbone}. Therefore, many experimental and theoretical 
investigations have been devoted to study the resistance noise as a function 
of temperature, bias strength and of the main material properties 
\cite{review,kogan,upon05,bosman,raychau_wire,soliveres,tarasov,tersoff,appenzeller,leturcq,torquato,sahimi,bardhan,grier,bloom,kakalios,lust94,alers,weissman_1f,planes_1f,carbone,chen_1f,raychaud,chiteme,em_fn04,ciliberto,sornette97,derrida,pre_fnl,pen_ng}. 
One of the most relevant features of the resistance noise lies in its  
dependence on frequency. Many condensed matter systems display the so called 
Lorentzian noise \cite{review,kogan}, which is characterized by a power
spectral 
density of the resistance fluctuations scaling as $1/f^2$ at high frequencies 
and becoming flat below a corner frequency $f_c$. A behavior associated 
in the time domain with an exponential decay of the correlations and thus with
a well defined characteristic time, $\tau$ (correlation time)
\cite{review,kogan}. 

On the other hand, it is well known \cite{review,kogan} that many other
condensed 
matter systems exhibit $1/f$ resistance noise, i.e. a spectral density scaling 
at low frequencies as $1/f^\alpha$ with $\alpha \approx 1$, thus a noise
associated with a non-exponential decay of the correlations in the time 
domain. The ubiquitous presence of $1/f$ noise in a large variety of phenomena
of very different nature has given rise to many attempts to explain it in 
terms of a universal law \cite{review,kogan,upon05}. Moreover the link
between $1/f$ noise and extreme value statistics has been also investigated 
\cite{racz1,racz2}. A simple way to obtain an $1/f$ spectrum is by 
superimposing a large number of Lorentzian spectra with an appropriate 
distribution of the correlation times \cite{review,kogan}. In some cases, this 
distribution can be derived from the distribution of some variable on which 
the correlation times themselves depend, as in the case of the pioneering 
works of Mc Whorter \cite{review,kogan} and Dutta et al. \cite{review,dutta79}. 
In particular, these authors proposed \cite{review,kogan,dutta79} that the
origin of the $1/f$ noise could be attributed to a thermally activated 
expression of the correlation times, associated with a broad distribution of 
the corresponding activation energies, an assumption physically plausible 
for many systems \cite{review,kogan,torquato,sahimi}.  
 
In the last twenty years, many other important contributions have advanced the
understanding of the $1/f$ noise showing that the presence of an $1/f$ spectrum
can also arise from other basic reasons  
\cite{upon05,grier,kakalios,zhang,weissman93,btw,zap_soc,jung,kaulakys_1f,kiss,celasco_1f,rakhmanov_1f,shklovski_1f,shtengel_1f,davidsen}. 
Thus, the conclusion, now largely accepted in the literature 
\cite{review,kogan,upon05,celasco_1f}, is that a unique, universal origin of 
$1/f$ does not exist, even though classes of systems can share a common basic 
origin of $1/f$ noise. For example, spin-glass models \cite{review,zhang,weissman93}
provide a good explanation of the $1/f$ noise in conducting random magnetic 
materials. Dissipative self-organised criticality (SOC) models \cite{btw,zap_soc} 
clarify the origin of $1/f$ spectra in certain dissipative dynamical systems 
naturally evolving into a critical state. Avalanche models \cite{jung}, 
clustering models \cite{kaulakys_1f} and percolative models 
\cite{torquato,sahimi,bardhan,grier,kakalios,kiss,celasco_1f} 
represent other relevant classes of theoretical approaches explaining 
the appearence of $1/f$ noise in a variety of systems.  
In particular, the use of random resistor network (RRN) models  
\cite{torquato,sahimi,stauffer,odor,arcangelis} has proved to be very 
fruitful. Within this approach, a large attention has been devoted to the 
calculation of the noise exponents in two-components RRN, as for example 
in Refs. \onlinecite{rammal85,park87,bergman,tremblay91}. 
However these authors assumed the existence {\em a priori} of 
independent $1/f$ microscopic fluctuators, associated with small 
fluctuations of the local resistivity, and  they studied the influence 
of the topology and disorder of the network on the resistance noise 
magnitude at a fixed frequency. The existence of microscopic 
Lorentzian fluctuators giving rise to $1/f$ noise at a macroscopic level, 
has been proposed instead by Gingl et al. \cite{gingl}. The hypothesis of 
small local fluctuations has been released by Seidler et al. \cite{seidler96}, 
who have shown that when the local resistivity fluctuations are large, the 
dynamical redistribution of the current gives rise to long-range (space) 
correlations and non-Gaussian $1/f$ noise. 

Finally, another important class of $1/f$ RRN models is represented 
by dynamical percolation models \cite{kakalios,lust94,celasco_1f}, 
introduced by Lust and Kakalios for describing Lorentzian spectra 
\cite{lust94} and then modified to account for $1/f$ noise 
\cite{kakalios,celasco_1f}. Within these models, $1/f$ noise 
arises from random jumps performed by some elemental component of the 
system between two states (like trapping and detrapping of charge carriers 
\cite{kakalios,lust94}, or ON and OFF states in the switcher model 
\cite{celasco_1f}). Precisely, Lust and Kakalios considered a two-dimensional 
RRN with half of the resistors removed at random (percolation threshold). 
By focusing on the filamentary resistive structures connecting 
the electrodes, they allowed fluctuations only for the resistors at the 
nearest neighbor positions \cite{kakalios,lust94}. In this manner, a uniform 
distribution of the trapping times provides a Lorentzian spectrum 
\cite{lust94}, while a certain non uniform distribution of these times gives 
rise to an $1/f$ noise spectrum \cite{kakalios}. Furthermore,  Celasco and 
Eggenh\"offner \cite{celasco_1f} have studied a binary RRN with links behaving
as random switchers of resistances $r$, fluctuating bewteen two ON and
OFF states (associated with $r=0$ and $r \neq 0$, respectively). The time 
evolution of a single switcher is controlled by two parameters, $p$ and $q$, 
respectively representing the probability for a switcher to be ON at the time 
$t$ and the probability for the same switcher to be off at $t+\Delta t$ if 
it was ON at $t$. This model applies only to linear systems and it does not 
contain any direct link with the external conditions, however its nice and 
nearly unique feature is that it provides a resistance noise with power 
spectrum $1/f^\alpha$ with $0 < \alpha < 2$ depending on the values of $p$ and 
$q$. Actually, many experiments report about $\alpha$ ranging between $0 \div 2$ 
in the same system depending on the external conditions 
\cite{review,kogan,upon05,bosman,alers,raychaud}. In particular, at high
external biases, transitions from $1/f$ to Lorentzian noise are observed in 
many systems, as a consequence of the suppression of the plurality of 
characteristic times induced by extreme, far from equilibrium 
conditions \cite{review,sornette}. 

The aim of this paper is to present a new model able to describe 
within a unified framework both Lorentzian and $1/f$ noises, together 
with the transitions between these two kinds of spectra \cite{review,kogan}. 
Thus, by connecting the differences in the power spectral density not only 
with the parameters intrinsic to a given system but also with the different 
external conditions: ranging from nearly equilibrium conditions (very low 
biases), to non-equilibrium stationary states, up to failure conditions. 
Correspondingly, in the time domain this means to tune the decay of 
correlations from a long-term decay (power-law scaling of the two-points 
auto-correlation function) up to an exponential decay.
 
The model that we propose belongs to the dynamical percolation 
class \cite{kakalios,lust94,celasco_1f}. 
Precisely, we consider a network made by several species of resistors, 
where each species is characterized by the values of the elementary resistance,
of the temperature coefficient and of two activation energies, which control 
the probabilities of breaking and recovery processes of that species. For this 
reason we call this model ``multi-species network'' (MSN) model. The states, 
either stationary or non-stationary, of the MSN result from the stochastic 
competition between the breaking and recovery processes of the different 
species. In analogy with Dutta et al. \cite{review,kogan,dutta79}, we take the 
activation energies distributed in a broad range of values, as discussed 
in the next section. As a consequence, the resistance fluctuations of the MSN 
exhibit an $1/f^{\alpha}$ power spectrum, where the value of $\alpha$ in 
the range $0 \div 2$ depends on temperature and current. 
 
The paper is organised as follows: in Sect. II we illustrate the MSN model, 
in Sect. III we report the results and finally in Sect. IV we draw the 
conclusions of this study.  
 
\section{Model} 
The multi-species network can be considered as a generalization of the single 
species network (SSN) \cite{pre_fnl,pen_ng,prl_fail,prl_stat,pen_prb,pen_ng_fn04}. 
The SSN model describes a RN whose resistance fluctuates over a single 
time-scale. In such a network the correlations relax exponentially (with small 
deviations from a single exponential decay at high biases) and the power 
spectral density of the resistance fluctuations is Lorentzian-like 
($\alpha \approx 2$) at all temperatures and currents 
\cite{pre_fnl,pen_prb,pen_ng_fn04}, as will be discussed in the following.
Both the models, MSN and SSN, share some technical features that the reader 
can found with more details in Refs. \cite{pre_fnl,prl_stat,pen_prb}. 

As usual for RRN models \cite{torquato,sahimi,stauffer,odor}, a 
conducting thin film with granular structure is described as a two-dimensional
$N \times N$ resistor network with square-lattice, where $N$ is the linear 
size of the network \cite{geo}. The RN is biased by an external constant 
current, $I$, applied through perfectly conducting bars placed at the left 
and right hand sides and it is in contact with a thermal bath at a 
temperature $T$. In the MSN model proposed here the network is made 
by $N_{spec}$ different species of resistors, whose resistances are 
denoted by $r_{n,i}$, where $n$ is the index of the elementary resistor 
specifying its position within the network and $i=1,...N_{spec}$ labels the 
species. The elementary resistances are taken linearly dependent on 
temperature: $r_{n,i} =r_{0,i}[1+\alpha_{T,i} (T_n - T)]$, where $r_{0,i}$, 
$\alpha_{T,i}$ are respectively the resistance at the thermal bath  
temperature and the temperature coefficient of the $i$-th species, while $T_n$ 
is the local temperature. By neglecting time dependent effects in the 
heat diffusion \cite{sornette92} we account for the local Joule heating
of the n-{\em th} resistor and its neighbors by expressing the local 
temperature as \cite{prl_fail}: $T_n = T + A \ \Delta_n$, 
where $A$ describes the heat coupling of the elementary resistor with the 
thermal bath and $\Delta_n$ depends on the local currents \cite{pre_fnl}.
  
Each resistor can be in two states: regular (with resistance $r_{n,i}$) or
broken (with resistance $r_{OP} = 10^9 r_{n,i}$). Resistors in the broken
state are called defects. The transition from the two states is
stochastic: the transition from the regular state to the broken one
(breaking process) occurs with probability $W_{D,i}$ while the reverse
transition (recovery process) occurs with probability $W_{R,i}$, where $i$
is the species index defined above. Both processes are thermally
activated, thus their probabilities are:  $W_{D,i}=\exp(-E_{D,i}/k_B T_n)$
and $W_{R,i}=\exp(-E_{R,i}/k_B T_n)$, where $E_{D,i}$ and $E_{R,i}$ are
the activation energies of the $i$-species and $k_B$ is the Boltzmann
constant. 

The initial state of the network (equal or different fraction of each 
species, ordered or randomly distributed within the network, values of 
$r_{0,i}$ and $\alpha_{T,i}$) has an important role on the network evolution: 
thus both on the actual achievement of stationary states and on the features 
of the resistance fluctuations. The initial conditions that we have considered 
in this work are the following. 

i) We have taken the values of $r_{0,i}$ and $\alpha_{T,i}$ uniformly 
distributed inside a given range of values: 
respectively $r_{0,i} \in [r_{min},r_{max}]$ and $\alpha_{T,i}
\in [\alpha_{min},\alpha_{max}]$. This choice has been adopted 
for the sake of simplicity, while other options are reasonable as well. 

ii) We have assumed a random distribution of the species within the network. 
Though special patterns for the space distribution of the species within 
the network can be of interest in many situations, the choice adopted here 
is the simplest one. 

iii) We have taken the activation energies $E_{D,i}$ and $E_{R,i}$ of the 
different species uniformly distributed inside the ranges of values 
$[E_D^{min},E_D^{max}]$ and $[E_R^{min},E_R^{max}]$.  This choice, 
taken in analogy with Dutta et al. \cite{review,kogan,dutta79}, 
is physically justified in many disordered systems, like granular and 
amorphous materials, composites, etc., where the orientational disorder 
present inside these materials can give rise to different energy barriers 
for the electron flow along the different conducting paths 
\cite{review,kogan,torquato,sahimi,zhang}.

vi) The energies, $E_{D,i}$, $E_{R,i}$ have been coupled by imposing the 
condition that the difference $E_{D,i}-E_{R,i}$ is approximately the same for 
the different species: $E_{D,i}-E_{R,i} \approx \Delta E^*$. The reasons of 
this assumption will be explained below. However, we anticipate that by 
defining $p_i$ as the fraction of defects belonging to the $i$-th species: 
$p_i \equiv N_{brok,i}/N_{tot,i}$, and $\tau_i$, as the correlation time which 
characterizes its fluctuations, the conditions iii) and vi) imply a 
logarithmic distribution of the correlation times of the different species.  

v) The association between the resistance $r_{0,i}$ of the $i$-th species and
the corresponding activation energies $E_{D,i}$ and $E_{R,i}$ has been done
by adopting the criterion that increasing values of $\tau_i$ are paired with
increasing values of $r_{0,i}$. Alternative choices are of course possible.

The physical meaning and the implications of the assumptions iii) and iv) 
can be understood by the following arguments, which also provide a 
theoretical ground to the numerical results reported in Section III.

Let us consider a network made by a single species of resistors at the
equilibrium, in the vanishing current limit $I  \rightarrow 0$ 
(Ref. \onlinecite{prl_stat}). In this case, 
it is easy to derive the following expression for the average fraction of 
defects under stationary fluctuations \cite{prl_stat}:
\begin{equation} 
\langle p \rangle = { W_D(1-W_R) \over  W_D(1-W_R) + W_R} =
{1 \over 1 + {W_R \over W_D(1-W_R)} } \label{eq:frac} 
\end{equation}
where $W_{D}$ and $W_{R}$ are the probabilities of the breaking and
recovery processes (for a SSN these quantities are independent of the
index $i$ and, at the equilibrium, they are also independent of the index $n$
specifying the position within the network). Now, it is convenient to 
introduce the following definition: 
$\lambda \equiv \ln(W_R/W_D)=(E_{D}-E_{R})/k_B T$. Since $W_R \ll 1$,  
Eq.~(\ref{eq:frac}) can be approximated as:
\begin{equation} 
\langle p \rangle \approx {1 \over 1 + e^{\lambda} }
\label{eq:lambda}
\end{equation}
Since $\lambda$ controls the average fraction of defects at the equilibrium, 
determining the ``intrinsic'' disorder of the network, it can be called  
intrinsic disorder parameter. Thus, the energy difference $E_{D}-E_{R}$ is 
the effective activation energy which sets the average defect 
fraction in the network. Furthermore, we note that a SSN can be considered as
a network subjected to random telegraph noise (RTN) of the elemental 
resistors, where each resistor $r$ fluctuates between two states, $1$ (active) 
and $2$ (broken), with the following transition rates: 
$W(1 \rightarrow 2 )=W_D$ and $W(2 \rightarrow 1)=W_R/(1-W_R)$. Thus, 
according to RTN theory \cite{kogan}, the power spectral density of the 
elemental fluctuator is given by:
\begin{equation}
S_{r}(f) ={ 4 \langle p \rangle (1-\langle p \rangle) \rho \tau_r 
\over 1 + (2\pi f \tau_r)^2} \label{eq:loren}
\end{equation}
where $\rho$ is the difference in the resistance of the two states and 
$\tau_r$ is the correlation time of the elemental fluctuator, given by:  
\begin{equation} 
{1\over \tau_r} = W(1 \rightarrow 2) + W(2 \rightarrow 1)= 
W_D + {W_R \over (1-W_R)}={W_D \over \langle p \rangle} \label{eq:tau}
\end{equation}
In Ref. \cite{prl_stat} it has been shown that this expression holds also for
the correlation time $\tau$ of the fluctuations of the global network 
resistance $R$, at least when the average fraction of defects 
$\langle p \rangle$ is sufficiently far from the percolation threshold 
\cite{stauffer,percol}. In other terms, under the last condition, it is: 
$\tau=\tau_r$. On the other hand, it is well known that when a global quantity,
like the network resistance, results from the superposition of several
exponential relaxation processes  with different correlation times, its
spectral density can be written as \cite{review,kogan}:
\begin{equation}
S_{R}(f) = \int_0^\infty d\tau'  g_R(\tau') {  4\tau' \over 1 + (2\pi f \tau')^2} 
\label{eq:spectrum}
\end{equation}
where $g_R(\tau')d\tau'$ specifies the contribution to the fluctuations of
the global quantity by the elemental processes whose correlation times
lie in the interval between $\tau'$ and $\tau' + d\tau'$. In conclusion, since
for a SSN at the equilibrium  $g_R(\tau')=\delta(\tau'-\tau)$, the power 
spectral density of the network resistance fluctuations takes the
Lorentzian form of Eq.~(\ref{eq:loren}).

The study of non equilibrium stationary fluctuations of a SSN  
\cite{pre_fnl,pen_prb} has shown that many of the effects of an external
current can be approximately described within a mean field-like framework, 
by considering average transition probabilities, $\langle W_D \rangle$ and 
$\langle W_R \rangle$  \cite{pen_prb}, which depend on the bias through the 
average temperature: $\langle T \rangle_I = T + \theta R_0 I^2$, where 
$\theta$ is the structure thermal resistance and $R_0$ is the network 
resistance in the vanishing current limit. Of course, at high external bias, 
the distribution of currents and temperatures within the network becomes 
strongly non homogeneous, resulting in breaking and recovery probabilities, 
$W_D=exp(-E_D/k_B T_n)$ and $W_R=exp(-E_R/k_B T_n)$, strongly dependent on 
the position of the elemental resistor within the network. This effect, which 
gives rise to a filamented growth of the defect pattern, characteristic of
biased percolation \cite{pre_fnl,pen_prb}, also implies different correlations 
times for the fluctuations of the elemental resistors. However, this 
difference in the correlation times, is never so large to modify significantly 
the shape of the power spectral density of the network resistance fluctuations,
which remains Lorentzian-like at all currents compatible with stationary
states of the network \cite{pre_fnl,pen_prb}. In other terms, the density 
function $g_R(\tau)$ in Eq.~(\ref{eq:spectrum}) is different from zero
only in a relatively small interval of $\tau$ values.  
 
Coming back to a MSN, it is convenient to define for all the species $i$ 
the quantities $\lambda_i \equiv (E_{D,i}-E_{R,i})/k_B T$, which control the 
average fraction of defects $\langle p_i \rangle$ at the equilibrium. Thus, 
the value of $\lambda_i$ determines the contribution of the $i$-th species
to the ``intrinsic'' disorder of the network. Then, by taking: 
$\lambda_i \approx cost =\Delta E^*/k_B T$, $\forall i$, we are assuming
an approximately equal concentration of the different species at the
equilibrium. In such situation, we expect that a mean field-like 
approach works and that the energy $\Delta E^*$ plays the same role of
effective activation energy as that played by the energy difference 
$E_D - E_R$ in the SSN at equilibrium. In other terms, $\Delta E^*$ controls 
the average defect fraction in the network. Furthermore, it must be noted 
that the conditions iii) and vi), coupled with Eqs.~(\ref{eq:lambda}) 
and ~(\ref{eq:tau}), imply a logarithmic distribution of the correlation 
times: $g(\tau_i)=1/\tau_i$, with $\tau_i \in [\tau_{min},\tau_{max}]$, 
where $\tau_{min}$ and $\tau_{max}$ define the time interval in which 
the correlation times are distributed. Since $W_{D,i}$ and $W_{R,i}$ depend 
also on $T_n$ (i.e. on the external temperature and on the local Joule 
heating) a logarithmic distribution of $\tau_i$ is obtained only at the 
equilibrium, in the vanishing current limit, when the local Joule heating is 
negligible, $\Delta_n \approx 0$, and all the resistors are at the same 
temperature: $T_n=T$. Moreover, apart from this effect related to a non 
homogeneous current distribution, in the nonlinear regime there is an 
increase of the average temperature which modifies the average transition 
probabilities, changing $\tau_{min}$ and $\tau_{max}$, according to 
Eq.~(\ref{eq:tau}). In  conclusion, the values of $\tau_{min}$ and 
$\tau_{max}$ depend on the particular material ($E_D^{min}$, $E_D^{max}$, 
$E_R^{min}$, $E_R^{max}$) and on the external conditions ($T$ and $I$).

The time evolution of the network is then obtained by Monte Carlo simulations 
which update the network resistance after a sweep of breaking and recovery 
processes, according to an iterative procedure detailed in 
Ref. \onlinecite{pre_fnl}. The sequence of successive network configurations 
provides a resistance signal, $R(t)$, after an appropriate calibration of the 
time scale. Then, depending on the external conditions and on the network 
parameters, the network either reaches a steady state or undergoes an 
irreversible electrical failure \cite{pre_fnl,pen_ng,pen_prb,pen_ng_fn04}.  
This latter possibility is associated with the condition that the global 
average defect fraction $\langle p \rangle = \sum_i \langle p_i \rangle$ 
reaches the percolation threshold, $p_c$.  
Therefore, for a given network at a given temperature, a threshold current 
value, $I_B$, exists above which electrical breakdown occurs \cite{pre_fnl}. 
For current $I \le I_B$, the steady state of the network is characterized by 
fluctuations of the defect fraction, $\delta p$, and of the resistance, 
$\delta R$, around their respective average values $\langle p \rangle$ and 
$\langle R \rangle$.
 
All the results reported here concern networks of sizes $75 \times 75$ made 
by $N_{spec}=15$ resistor species. For the other parameters the following 
values have been used as representative of realistic cases: $r_{min}=0.5$ 
$\Omega$ and $r_{max}=1.5$  $\Omega$, $\alpha_{min}=10^{-4}$ K$^{-1}$, 
$\alpha_{max}=10^{-1}$ K$^{-1}$.
Moreover, we have taken: $E_D^{min}=58$ meV, $E_D^{max}=375$ meV, 
$E_R^{min}=37$ meV, $E_R^{max}=346$ meV and $\Delta E^* \approx 25$ meV 
(precisely, the average value of the difference between $E_{D,i}/k_B$ and 
$E_{R,i}/k_B$ is $\Delta E^*/k_B \approx 319.86$ K). 
 
According to Eq.~(\ref{eq:tau}) and conditions iii) and iv), in
equilibrium (or nearly equilibrium) at the reference temperature
$T_{ref}=300$ K, the above energy values imply 
$\langle p_i \rangle \approx 0.25 \  \ \forall i$, $\tau_{min} \approx 2$ 
and $\tau_{max} \approx 5 \times 10^5$ (where times are expressed in units 
of iterative steps). At $T > T_{ref}$, the interval defined by $\tau_{min}$
and $\tau_{max}$ becomes progressively narrower while $\langle p_i \rangle $ 
increases. Of course, the contrary happens for $T<T_{ref}$. We underline that 
these values of $\langle p_i \rangle$, $\tau_{min}$ and $\tau_{max}$, 
calculated from  Eq.~(\ref{eq:tau}) are reported here just to give a 
qualitative idea of the network state and because they can be useful for the 
choice of the parameters $E_D^{min}, E_D^{max}, E_R^{min}, E_R^{max}$. 
However all the network properties, including the average defect fraction, 
the correlation time of the resistance fluctuations and the other results 
discussed in Sec. III are obtained directly from the output of simulations. 
Finally, the auto-correlation functions and the power spectral densities 
of the resistance fluctuations are calculated by analyzing  
stationary $R(t)$ signals consisting of $1 \div 2 \times 10^6$ records. 

\section{Results} 
Figure 1 reports the resistance evolution of a MSN calculated at $300$ K in the
vanishing current limit. The inset displays a small part of the same evolution
over an enlarged time scale. Here we notice the co-existence of different 
characteristic time scales in the $R(t)$ signal. Indeed, the long relaxation 
time associated with the achievement of the steady state, $\tau_{rel}$, 
co-exists with the shorter times characterizing the resistance fluctuations, 
as displayed in the inset. For comparison, Fig. 2 reports the time evolution 
of the resistance of a SSN obtained at $T=300$ K in the same bias conditions. 
In this case, the values of the activation energies ($E_D=350$ meV and 
$E_R=310$ meV) are chosen to give a relaxation time comparable with that of 
the signal in Fig. 1. Now, the resistance signal is controlled by a single 
time scale ($\tau \approx \tau_{rel}$) and it is essentially flat over time 
scales shorter than $\tau$. This is emphasized by the inset in Fig. 2 where 
the stochastic signal resembles that of a few levels system (it must be noted 
that the vertical scale of the inset in Fig. 2 is significantly enhanced with 
respect to that of Fig. 1).  
 
Figure 3 displays the auto-correlation functions of the resistance fluctuations
in Figs. 1 and 2, corresponding to the MSN model (black triangles) and to the 
SSN model (black squares). A log-log representation is adopted for 
convenience. The solid and short dashed lines represent the best-fits to the 
two auto-correlation functions carried out, respectively, with a power-law and
an exponential law. The fitting procedure confirms the exponential decay of 
the correlations in the resistance fluctuations of the SSN. Furthermore, 
it points out the existence of long-term correlations in the resistance
fluctuations of the MSN, characterized by a power-law decay of the 
auto-correlation function: 
\begin{equation} 
C_{\delta R}(t) \sim  t^{-\gamma}  \label{eq:power_law} 
\end{equation} 
with $0 < \gamma <1$. In particular, here we have found a value 
$\gamma=0.22 \pm 0.01$ for the correlation exponent. We stress that the 
above expression for the auto-correlation function implies a divergence of 
the correlation time, as can be easily seen by considering the following 
general definition of the correlation time \cite{bunde_pa2003}: 
\begin{equation} 
\tau = \int_{0}^{\infty} {C_{\delta R}(t) \over C_{\delta R}(0)}dt \label{eq:tau_def} 
\end{equation} 
Figure 4 shows the power spectral densities of the resistance fluctuations
of a MSN and of a SSN calculated at 300 K by Fourier trasforming the
auto-correlations functions of Fig. 3. Here, the two grey solid lines 
represent the best-fits with a power-law of the MSN spectrum 
in the low and high frequency regions. We can see that the resistance 
fluctuations of the multi-species network exhibit at low frequencies a power 
spectral density scaling as $1/f^\alpha$, with a value $\alpha=0.94$. We 
notice that this scaling behavior holds over several decades of frequency. 
This result is a consequence of the envelope of the different time scales 
associated with the different resistor species, described by 
Eq.~(\ref{eq:spectrum}). Instead, in the high frequency region, the slope
of the spectrum is -1.53 (in fact, the slow relaxations are ineffective 
at such high frequencies). For contrast, the grey dashed curve in Fig. 4 
is the best-fit with a Lorentzian to the SSN spectrum. The corner frequency 
of the Lorentzian, $f_c=4.0 \times 10^{-6}$ (arbitrary units), is consistent 
with the correlation time reported in Fig. 3 and obtained by the best-fit of 
the corresponding auto-correlation function.
 
Now, we will discuss how the temperature of the thermal bath affects the 
resistance fluctuations of a multi-species network at equilibrium or nearly 
equilibrium conditions (low biases). In other terms, we will analyse the
properties of the fluctuations in the Ohmic regime and for different 
temperatures. Figure 5 reports the resistance evolutions calculated at 
increasing temperatures: $T=400$ K (lower curve) and $T=600$ K (upper curve). 
Already this qualitative comparison between the $R(t)$ signals shows 
that a temperature increase implies a significant growth of both the average 
resistance and the variance of the resistance fluctuations (see also Figs. 1). 
Furthermore, this comparison points out a drastic reduction of the relaxation 
time at increasing temperatures (more than one order of magnitude when $T$ 
rises from 300 K to 400 K).  
 
The temperature is also found to affect the distribution of the resistance  
fluctuations by increasing its skewness, as shown in Fig. 6, which reports the 
probability density function (PDF) of the resistance fluctuations for several 
temperatures. A normalized lin-log representation has been adopted here
for convenience ($\sigma$ is the root mean square deviation from the average 
resistance). The normalized Gaussian distribution is also reported for
comparison. The figure shows a significant non-Gaussianity of the resistance 
fluctuations, which becomes stronger at high temperatures, when the system
approaches failure conditions. This behavior is completely different from
the behavior of the defect fraction fluctuations, shown in Fig. 7, which
remains Gaussian at all temperatures. This different behavior can be
easily understood. Actually, at increasing temperature, the local resistivity 
fluctuations become progressively larger. This fact emphasizes the dynamical 
redistribution of currents and brings to the emergence of long-range
correlations inside the network \cite{seidler96}. Then, the violation of the
validity conditions of the central limit theorem leads to a non-Gaussian
resistance noise \cite{review,kogan,weissman_1f,seidler96,ausloos}. 
In other terms, the results in Fig. 6 imply that the 
correlation length of the resistance fluctuations progressively increases 
with the temperature. A discussion of the distribution of the resistance 
fluctuations and the role played on the non-Gaussianity by the size, shape 
and disorder of a SSN is reported in Refs. \onlinecite{pen_ng,pen_ng_fn04}, 
where the link with the universal distribution of the fluctuations of 
Bramwell, Holdsworth and Pinton (BHP) \cite{bramwell,clusel}  is analyzed. 
  
Figure 8 reports the auto-correlation functions of the resistance
fluctuations of a MSN calculated at $400$ K and $600$ K. To help the 
comparison, the auto-correlation function at $300$ K (already shown in Fig. 3)
has been drawn again in Fig. 8, together with its power-law best-fit 
(grey solid line). The dashed grey curves represent the best-fit to the
correlation functions at $400$ K and $600$ K with the expression: 
\begin{equation} 
C_{\delta R} (t)=C_0t^{-h} \exp(-t/u)  \label{eq:power_exp} 
\end{equation} 
The values of the best-fit parameters are: $C_0=1.13$, $h=0.30$ and  
$u=1.42 \times 10^4$ for $T = 400$ K and $C'_0=0.965$, $h'=0.46$ and  
$u'=5.57 \times 10^2$ for $T = 600$ K. The fit to $C_{\delta R}$ with 
Eq.~(\ref{eq:power_exp}) is found to be very satisfactory. We conclude 
that at $T > 300$ K, the auto-correlation function of the resistance 
fluctuations of the MSN is well described by a power-law with an exponential 
cut-off. Such kinds of mixed decays of the correlations, non-exponential and 
non power-law, are often found in the transition of a complex system from
short-term correlated to long-term correlated regimes
\cite{sornette,weissman93,franz03,franz06,pen_epj,caccioli,coniglio,parisi}. 
It should be noted that for $u \rightarrow \infty$, Eq.~(\ref{eq:power_exp}) 
becomes a power-law, while for $h \rightarrow 0$, it describes an exponential 
decay. Actually, Fig. 8 highlights a strong reduction of the correlation time 
of the resistance fluctuations as the temperature increases. 
This result can be understood in terms of Eq.~(\ref{eq:tau}), 
by considering the thermally activated expressions of the 
breaking and recovery probabilities. Indeed, the increase of temperature above
$T_{ref}$ implies the reduction of the ratio $\tau_{min}/\tau_{max}$ and
the correspondent narrowing of the interval $[\tau_{min},\tau_{max}]$, 
where the $\tau_i$ are distributed. This trend tends to suppress the 
power-law decay of correlations in favor of the exponential decay.  
For similar reasons, the temperature decrease below $T_{ref}$, implies
the opposite trend with the correlations keeping their power-law decay over 
wider time scales. For example, at $T=200$ K the ratio $\tau_{min}/\tau_{max}$
becomes  $7.5 \times 10^7$. Thus, we can define the transition
temperature, $T^*$, as the temperature value which signs the crossing from
a long-term correlated behavior (occurring for $T<T^*$), to a behavior 
characterized by a finite and relatively short correlation time 
(occurring for $T>T^*$).    
 
The correlation time of the resistance fluctuations can be directly 
estimated by making use of Eq.~(\ref{eq:tau_def}) and 
Eq.~(\ref{eq:power_exp}). In terms of the best-fit parameters of the 
auto-correlation function, it is easy to derive the following analytical 
expression for $\tau$: 
\begin{equation} 
\tau = u^{1-h} \Gamma(1-h) \label{eq:tau_fit} 
\end{equation} 
where $\Gamma$ is the Gamma function. The values of $\tau$ calculated in this
way increase at decreasing temperatures. In particular, $\tau$ exhibits a
sharp increase when $T$ approaches $T^*$, in agreement with the
long-term decay of the correlation function at this temperature. We have found
that the behavior of $\tau$ versus temperature is well fitted by the 
power-law: $\tau \sim (T-T^*)^{-\theta}$. Then, the fit procedure allows
us to determine the value of the transition temperature. We have found:
$T^*=306$ K and $\theta=2.7$. Figure 9 reports the values of $\tau$ as a 
function of the difference $T-T^*$. The dashed straight line corresponds to 
the above mentioned power-law. Therefore, we can conclude that: 
$T^* \approx \Delta E^*/k_B$. Furthermore, as a result of the particular 
choice of the parameters adopted in the present calculations, the transition 
temperature is close also to the reference temperature, $T^* \approx T_{ref}$. 
However we remark that the temperature $T_{ref}$ has been introduced in
the model merely to help the choice of the activation energies (to
provide a sufficiently wide interval $[\tau_{min},\tau_{max}]$). While,
$\Delta E^*$ is the relevant input parameter which determines  $T^*$.
  
Figure 10 displays the spectral densities of the resistance fluctuations 
calculated at $400$ K and  $600$ K. The grey lines represent the best-fits 
with a power-law of the MSN spectra in the low and high frequency regions.
Precisely, at low-frequencies, the slopes of the lines are -0.87 and -0.78  
respectively for $T=400$ K and $T=600$ K. Instead, at high frequencies the 
respective slopes are -1.31 and -1.02. Thus, in both frequency regions, 
low and high, the slopes of the specra are reduced when the temperature 
increases. We conclude that at $T>T^*$ the power spectrum keeps the 
$1/f^{\alpha}$ form with the value of ${\alpha}$ significantly decreasing
below unity. This decrease is pointed out in Fig. 11, which reports ${\alpha}$
as a function of the temperature. The dashed line in Fig. 11 is the best-fit 
with a linear law. We notice that such a decrease of the exponent, from 
$\alpha \approx 1$ to $\alpha \approx 0.8-0.5$, is frequently observed in 
the experiments at intermediate temperatures
\cite{review,kogan,upon05,bosman,alers}. 
On the other hand, many experiments have pointed out a strong dependence of 
the detailed behavior of $\alpha(T)$ also on the particular material 
\cite{review,kogan,upon05,alers}. In this respect, we remark that in this work 
we are focusing our attention to the general features of the spectra, related
to the term of correlations, rather than to the interpretation of a particular 
set of experiments. Actually, in our model the quantity  $\Delta E^*$ 
(determining the transition temperature $T^*$) is the input parameter which 
can be adjusted for a quantitative fit of experiments. Furthermore, we stress 
that the monotonic decrease of ${\alpha}$ for $T>T^*$ is obtained here in the 
linear regime of currents, i.e. neglecting Joule heating effects. 
Actually, these effects, whose importance also depends on the
temperature, can give rise to a more complicated behavior of ${\alpha}$ 
versus $T$, which are outside the interest of this research but can be 
implemented with minor efforts. 
 
Now, to complete the analysis of the MSN model, we consider the dependence on 
the temperature of the average resistance, $\langle R \rangle$, and of the
variance of the resistance fluctuations, $\langle (\Delta R)^2 \rangle$ 
for a multi-species network in nearly equilibrium conditions. This dependence 
has been already qualitatively depicted in Figs. 1 and 5. However here we want 
to quantify these behaviors, also with the purpose of checking by numerical
results the validity of the mean field-like approach discussed in Section II. 
Figure 12 displays $\langle R \rangle$ as a function of the temperature,
while the inset shows the dependence on the temperature of the average defect 
fraction. We have performed a one parameter best-fit of the numerical
values of $\langle p \rangle$ with the expression:  
$\langle p \rangle =c_p/( 1 + e^{\Delta E^*/K_BT})$. We have found 
$c_p=0.93 \pm 0.03$, a value compatible with the expression: 
$\langle p \rangle \approx 1 /(1 + e^{\Delta E^*/K_BT})$. Thus, 
Eq.~(\ref{eq:lambda}) qualitatively accounts for the behavior of the
average defect fraction versus temperature.

For what concerns the dependence on temperature of the average resistance,
in percolation theory \cite{stauffer} it is well known the following scaling
relation between the network resistance and the defect fraction: 
$R \sim |p - p_c|^{-\mu}$. We note that in the present model at the
vanishing current limit \cite{prl_stat}, the percolation is uncorrelated
and the exponent $\mu$ takes the universal value \cite{stauffer} 
$\mu=1.303$, while $p_c=0.5$ \cite{percol,stauffer}. 
The dashed curve in Fig. 12 shows the best-fit of the numerical data 
for the average resistance with the expression:
$\langle R \rangle=c_R |\langle p \rangle - p_c|^{-\mu}$. In the best-fit
procedure the values of $\mu$ and $p_c$ have been taken fixed to their
theoretical values, while we have taken: 
$\langle p \rangle =c_p/( 1 + e^{\Delta E^*/K_BT})$. 
The values found for the fitting parameters are: 
$c_R=0.42 \pm 0.01$ $\Omega$ and $c_p=0.80 \pm 0.03$. 
Thus, the dependence on temperature of the average resistance of a MSN in 
the linear regime is well described by the usual scaling relation 
coupled with the mean field-like expression of $\langle p \rangle$. 

Figure 13 reports the variance of the resistance fluctuations as a function 
of the temperature. This behavior can be easily understood by considering 
before the power-law relation between the variance of the resistance 
fluctuations and the average  resistance \cite{stauffer,rammal85}: 
$\langle (\Delta R)^2 \rangle \sim \langle R \rangle^{\eta}$, often
written in terms of the relative variance, as:
\begin{equation} 
\langle (\Delta R)^2 \rangle /\langle R \rangle^2 \ \sim  
\langle R \rangle^s \label{eq:relvar} 
\end{equation} 
where $s=\eta - 2$. By reporting on a log-log plot the relative variance 
$\langle (\Delta R)^2 \rangle/\langle R \rangle^2$ versus $\langle R \rangle$,
the calculated slope provides for the exponent $s$ the value $s=2.6 \pm0.01$ 
We note that this value of the relative noise exponent agrees with the value
reported in Ref. \onlinecite{prl_stat}. This agreement is consistent with
the fact that the MSN model generalizes the results of Ref. 
\onlinecite{prl_stat} to networks characterized by $1/f^{\alpha}$ noise.  
Now, we can use the information on the exponent $\eta=s+2=4.6$ to check
the consistency among the dependence on temperature of 
$\langle (\Delta R)^2 \rangle$, Eq.~(\ref{eq:lambda}), the scaling
relations between $\langle R \rangle \sim \langle p \rangle$ and between 
$\langle (\Delta R)^2 \rangle \sim \langle R \rangle$.
The solid curve in Fig. 13 shows the best-fit to the variance of the resistance
fluctuations considered versus temperature, with the following expression: 
$\langle(\Delta R)^2\rangle = c_\Delta c_R^{\eta} |\langle p \rangle-p_c|^{-\mu\eta}$ where the fitting parameters take the values: 
$c_\Delta=(4.42 \pm 0.04) \times 10^6$, $c_R=1.0 \pm 0.04$ and  
$c_p=0.74 \pm 0.03$. Thus, a mean field-like framework is overall able
to account for the dependence on temperature of the lowest two moments of
resistance fluctuation distribution.  
 
Until now we have analysed the resistance fluctuations of the network in
equilibrium or in nearly equilibrium conditions, in the Ohmic regime, and 
for different temperatures. Now, we will briefly discuss the non equilibrium 
properties of the fluctuations at high biases, in the nonlinear regime, and 
for a given temperature. Figure 15 shows the auto-correlation functions of the 
resistance fluctuations of a MSN for increasing values of the external current.
All the functions are calculated at $300$ K. The open squares  
represent the auto-correlation of the resistance fluctuations in the
Ohmic regime: $I=5$ mA, the other three curves correspond to the non-linear
regime. Precisely, black triangles:  $I=200$ mA; open circles: $I=220$ mA; 
grey diamonds: $I=250$ mA $>I_B$. Thus, the last curve corresponds to failure 
of the network, i.e. non-stationary resistance fluctuations. In this case, the
auto-correlation function has been calculated by considering only the 
nearly-stationary portion of the $R(t)$ signal (after subtraction of a 
linear trend). The dotted grey line is the best-fit with a power-law of 
slope $-0.22 \pm 0.01$, the dashed lines are the best-fits with
Eq.~(\ref{eq:power_exp}). The values of the parameter $h$
corresponding respectively to: $I=200$, $220$, $250$ mA are the following: 
$h=0.19$, $0.17$, $0.11$ (in all cases, the numerical error is estimated 
as $\pm 0.01$). The values of the parameter $u$ corresponding to the 
same currents respectively are:
$(1.26 \pm 0.01)\times 10^5$,  $(9.15 \pm 0.01)\times 10^4$, 
$(5.84 \pm 0.01)\times 10^3$. Thus, both $h$ and $u$ decrease at increasing
biases and for high biases $h \rightarrow 0$. In other terms, at increasing 
currents, time correlations decay progressively faster, the long-term
correlated behavior is destroyed and it emerges a trend towards a simple 
exponential decay. 

The corresponding effect in the frequency domain of increasing biases 
is shown in Fig. 16, which reports the power spectral densities at $300$ K.
The curve (1) is obtained for the Ohmic regime ($I=5$ mA); the other
three curves display the spectra in the non-linear regime. The solid grey 
lines are the best-fits with power-laws in the low and high frequency
regions of the spectrum. The slopes are specified in the figure. 
We note that: i) the corner frequency between the two regions progressively
moves towards lower frequencies at increasing biases; ii) the slopes in
the high frequency region are: $-1.56$, $-1.62$, $-1.66$, $-1.78$ respectively
for $I=5$, $200$, $220$ and $250$ mA, i.e. the slopes increase at increasing
currents (in all cases, the numerical error on the slope values is estimated 
as $\pm 0.01$). Thus, at high biases, the spectral densities show a trend 
towards a transition from an $1/f$ to a Lorentzian behavior. However, 
we remark that for the present choice of the initial conditions and/or 
the numerical values of the parameters used in these calculations we have 
not got a pure Lorentzian spectrum. 

\section{Conclusions}
We have developed a stochastic model to investigate the $1/f^{\alpha}$, with 
$0 < \alpha <2$, resistance noise in disordered materials. More precisely, 
we have considered the resistance fluctuations of a thin resistor with 
granular structure in different stationary states: from nearly equilibrium 
up to far from equilibrium conditions. Furthermore we have also considered 
fluctuations in non-stationary states, associated with failure of the
electric properties. This system has been modeled as a two-dimensional 
network made by different species of elementary resistors. The steady state 
of this multi-species network is determined by the competition among different 
thermally activated and stochastic processes of breaking and recovery of the 
elementary resistors. The network properties have been studied by 
Monte Carlo simulations as a function of the temperature and applied
current, in both Ohmic and non-Ohmic regimes. A mean field-like framework
has been also used to qualitatively describe the dependence on temperature 
of the lowest two moments of the resistance fluctuation distribution. 
Furthermore, the correlation properties of the resistance fluctuations have 
been analyzed in both the time and the frequency domains. The model gives 
rise to resistance fluctuations with different power spectra, depending on the 
external conditions. Thus it provides a unified approach to the study of 
materials exhibiting either Lorentzian noise or $1/f^{\alpha}$ noise. 
By analyzing the correlations in the time domain, it has been found that
the resistance fluctuations display a cross-over from long-term
correlations to intermediate-term correlations. Although a trend
towards an exponential decay has been identified, for the present choice
of the parameters/initial conditions, we have not got a single exponential
decay. However, the model proposed seems able to account for the complex
interplay exerted on the correlation properties of the resistance
fluctuations by the external conditions and resulting in the complicate
behavior of the noise exponent observed in many experiments.
In perspective, it should be explored the role of different initial
conditions: such as unequal presence of the species and/or non-uniform
distribution inside the network, which can give rise to
interesting spatio-temporal organization patterns.



\eject\vfill

\begin{figure}[bth]
\includegraphics[height=.25\textheight]{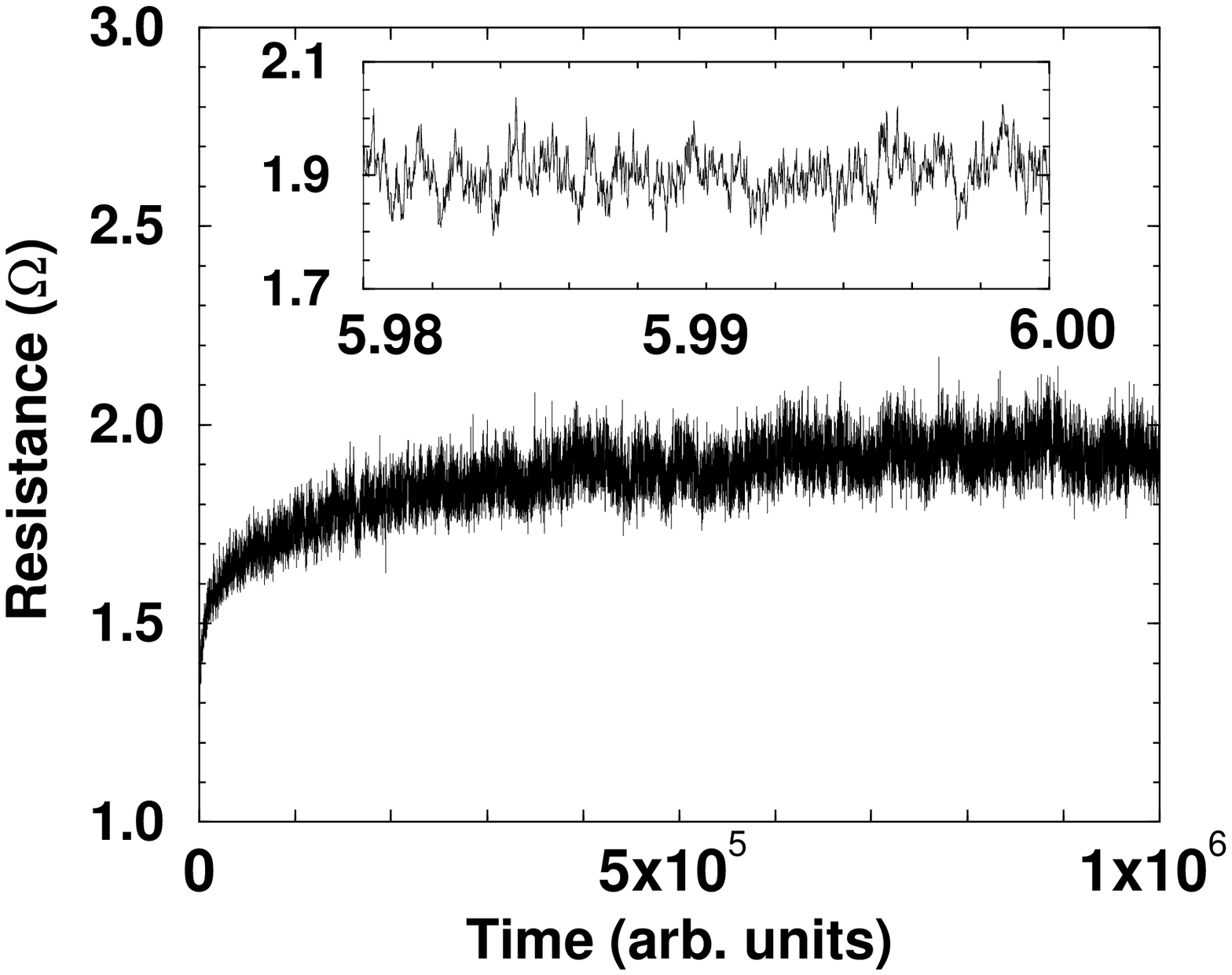}
\vspace*{0.2cm}
\caption{\label{fig:1}
Resistance evolution of a multi-species network (MSN model)
calculated at 300 K. The resistance is espressed in Ohm and the time
in iterative steps. The inset highlights the resistance fluctuations
on an enlarged time scale. In particular, the time units are divided   
by a factor $10^{-5}$.}
\end{figure}

\begin{figure}[bth]
\includegraphics[height=.25\textheight]{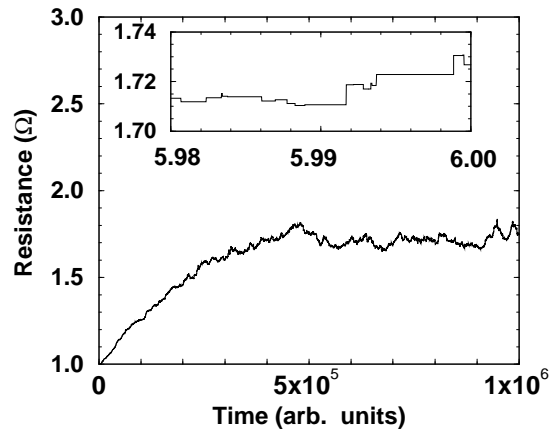}
\vspace*{0.2cm} 
\caption{\label{fig:2}
Resistance evolution of a single-species network (SSN model)
calculated at 300 K. The resistance is espressed in Ohm and the time
in iterative steps. The inset displays the resistance fluctuations on
the same enlarged time scale of the inset in Fig. 1. (the vertical  
scales of the insets in the two figures are different).}
\end{figure}

\begin{figure}[bth]
\includegraphics[height=.25\textheight]{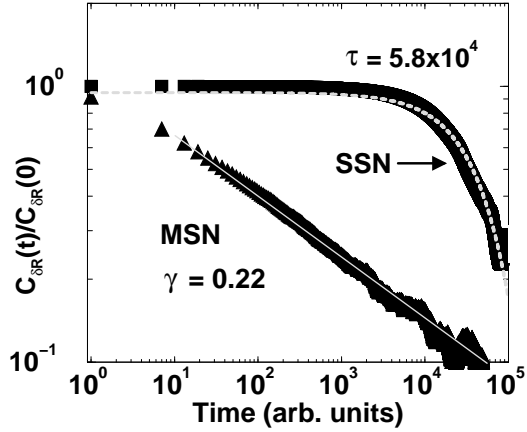}
\vspace*{0.2cm}
\caption{\label{fig:3}
Auto-correlation functions of the resistance fluctuations
calculated for a MSN (black triangles) and for a SSN (black squares).
Both functions are obtained at 300 K. The solid and short dashed grey
lines show the best-fit respectively with a power-law of exponent
$\gamma=0.22$ and with an exponential with correlation time 
$\tau=5.8 \times 10^4$. The time is expressed in iterative steps.}
\end{figure} 
\begin{figure}[bth]
\includegraphics[height=.25\textheight]{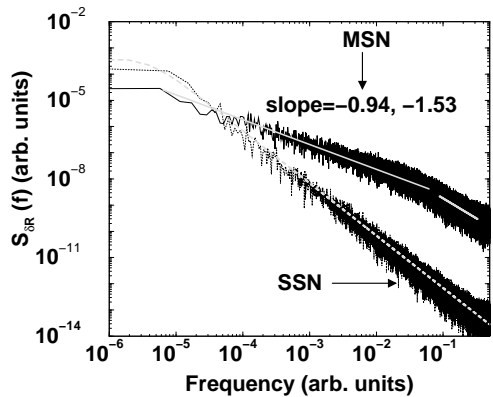}
\vspace*{0.2cm}
\caption{\label{fig:4}
Power spectral density of the resistance fluctuations at 300 K
calculated for a MSN (solid line) and for a SSN (dotted line). The grey
solid line shows the best-fit to the MSN spectrum with a power-law of
slope -0.94. The grey dashed curve represents the best-fit with a
Lorentzian to the SSN spectrum.}
\end{figure} 

\begin{figure}[bth]
\includegraphics[height=.25\textheight]{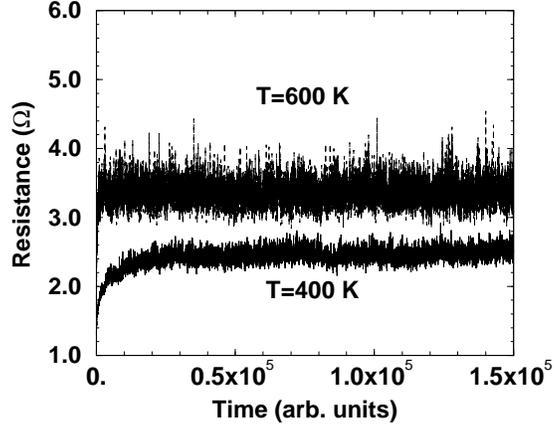}
\vspace*{0.2cm}
\caption{\label{fig:5}
Resistance evolution of a MSN at 400 K and 600 K.
The resistance is espressed in Ohm and the time in iterative steps.}
\end{figure}

\begin{figure}[bth]
\includegraphics[height=.25\textheight]{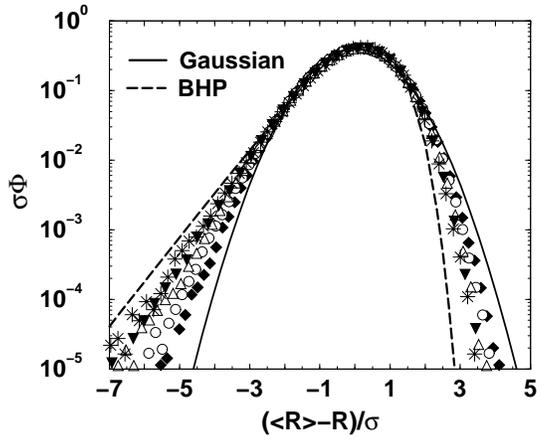}
\vspace*{0.2cm}
\caption{\label{fig:6}
Normalized probability densities of the resistance
fluctuations of a MSN calculated at 300 K (full diamonds), 400 K
(open circles), 500 K (open up-triangles), 600 K (full down-triangles)
and 700 K (stars). $\sigma$ is the root-mean-square deviation from the
average resistance. The solid black curve is the Gaussian
distribution and the dashed one the BHP distribution (see text).}
\end{figure}
 
\begin{figure}[bth]
\includegraphics[height=.25\textheight]{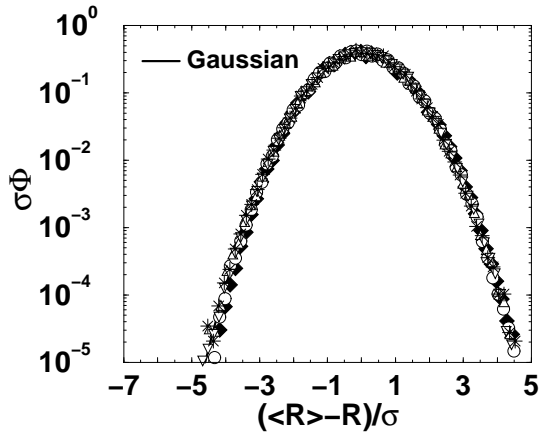}
\vspace*{0.2cm}
\caption{\label{fig:7}
Normalized probability densities of the defect fraction
fluctuations calculated at 300 K (full diamonds), 400 K
(open circles), 500 K (open up-triangles), 600 K (full down-triangles)
and 700 K (stars). Here $\sigma$ is the root-mean-square deviation from
the average value of the defect fraction. The solid black curve is the
Gaussian distribution.}
\end{figure}

\begin{figure}[bth]
\includegraphics[height=.25\textheight]{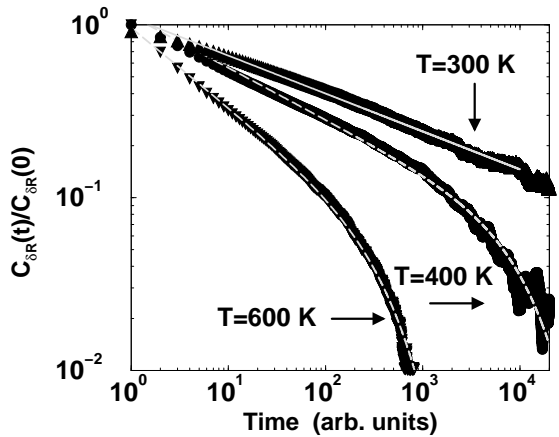}
\vspace*{0.2cm}
\caption{\label{fig:8}
Auto-correlation functions of the resistance fluctuations of
a MSN calculated at different temperatures. The solid grey
curve shows the best-fit with a power-law to the auto-correlation function
at 300 K (the same of Fig. 3). The dashed grey curves display the best-fit
to the auto-correlation functions at 400 and 600 K with the function:
$C(t)=C_0t^{-h} \exp[-t/u]$ (see the text for the values of the fit
parameters). The time is expressed in iterative steps.}
\end{figure}

\begin{figure}[bth]
\includegraphics[height=.25\textheight]{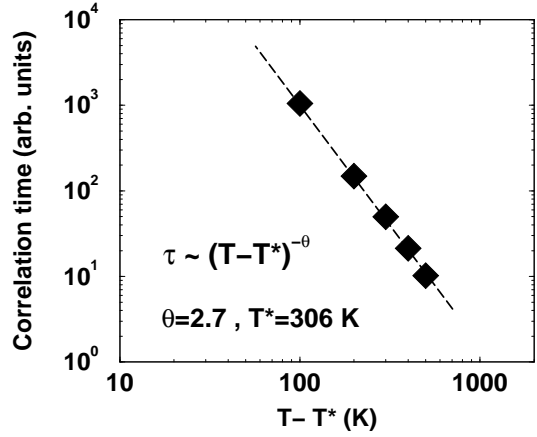}
\vspace*{0.2cm}
\caption{\label{fig:9}
Correlation time of the resistance fluctuations of a MSN as a 
function of the difference $T-T^*$. The time is expressed in iterative
steps and the temperature in K. The value of $T^*$ is reported in the
figure. The dashed line shows the fit with a power-law of exponent 
$\theta=2.7$.}
\end{figure}
 
\begin{figure}[bth]
\includegraphics[height=.25\textheight]{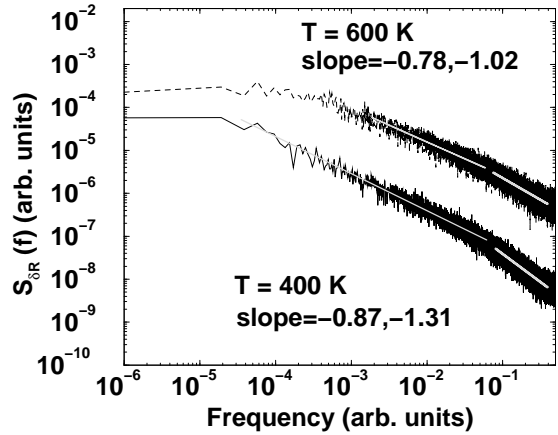}
\vspace*{0.2cm} 
\caption{\label{fig:10}
Power spectral density of the resistance fluctuations of a MSN
at $T=400$ K and $T=600$ K. The grey lines show the best-fit with
power-laws of slopes -0.87 and -0.78, respectively. The spectral density
at 600 K has been multiplied by a factor 10 for visual reasons.}
\end{figure}

\begin{figure}[bth]
\includegraphics[height=.25\textheight]{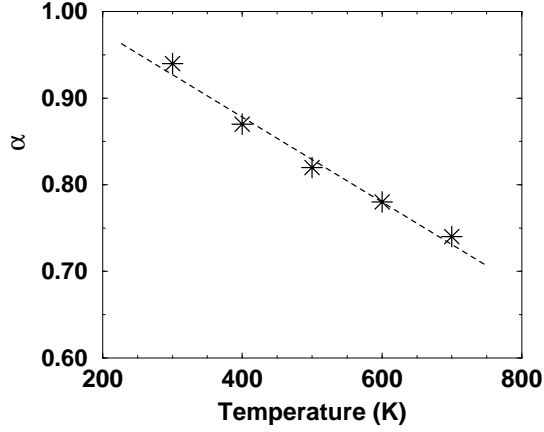}
\vspace*{0.2cm}
\caption{\label{fig:11}
Noise exponent $\alpha$ as function of the temperature. The dashed 
line refers to a linear best-fit of the calculated values of the exponent.}
\end{figure}

\begin{figure}[bth]
\includegraphics[height=.25\textheight]{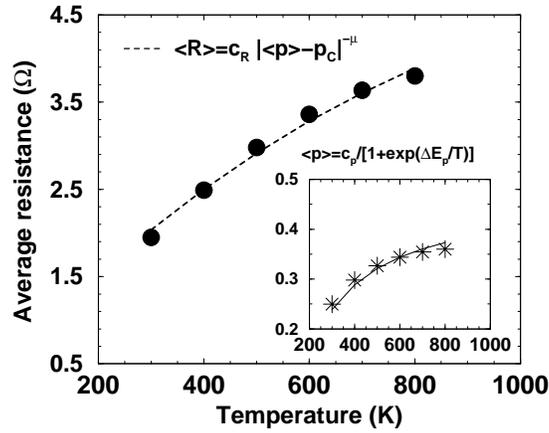}
\vspace*{0.2cm}    
\caption{\label{fig:12}
Variation of the average resistance (full circles) and average defect
fraction (stars, in the inset) as function of the temperature. The
resistance is espressed in Ohm and the temperature in K. The solid and
dashed straight lines are the best-fits with the expressions reported in
the figure (see the text for the values of fit parameters).}
\end{figure}

\begin{figure}[bth]
\includegraphics[height=.25\textheight]{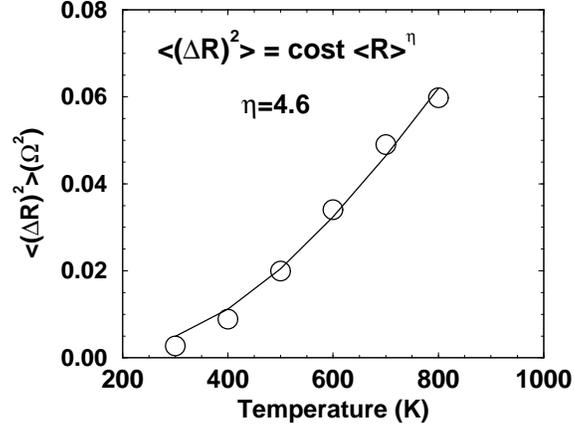}         
\vspace*{0.2cm}
\caption{\label{fig:13}
Variance of the resistance fluctuations as function of the
temperature. The variance is espressed in $\Omega^2$ and the temperature
in K. The dashed straight line shows a best-fit with the expression
reported in the figure (see the text for the values of fit parameters).}
\end{figure}  

\begin{figure}[bth]
\includegraphics[height=.25\textheight]{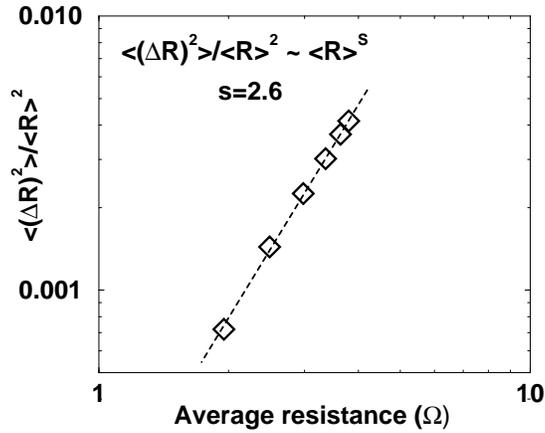}
\vspace*{0.2cm}
\caption{\label{fig:14}
Relative variance of the resistance fluctuations as a function of
the average resistance (this last expressed in Ohm). The dashed line shows
a best-fit with a power-law of exponent $s=2.6$.}
\end{figure}  

\begin{figure}[bth]
\includegraphics[height=.25\textheight]{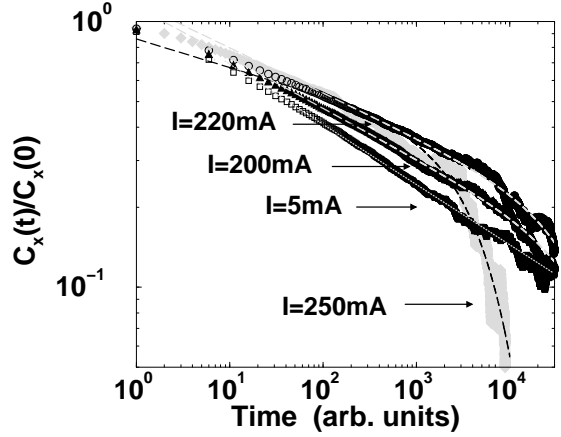}
\caption{\label{fig:15}
Auto-correlation functions of the resistance fluctuations of a
MSN at room temperature for increasing value of the external current.
Ohmic regime: $I=5$ mA (open square), non-linear-regime: $I=200$ mA
(black triangles), $I=220$ mA (open circles) and $I=250$ mA (grey
diamonds). The dotted grey line is the best-fit with a power-law, the
dashed lines are the best-fit with the function:  $C(t)=C_0t^{-h}
\exp[-t/u]$ (see the text for the values of the fit parameters).}
\end{figure}

\begin{figure}[bth]
\includegraphics[height=.25\textheight]{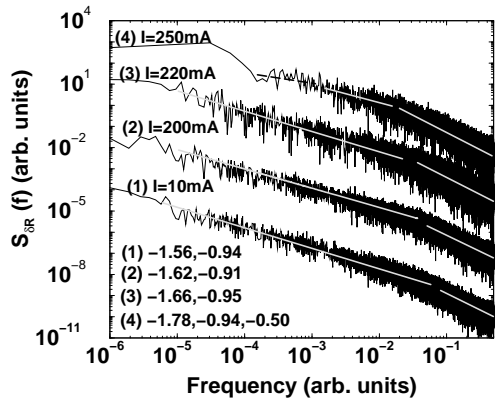}
\vspace*{0.2cm}
\caption{\label{fig:16}
Power spectral density of the resistance fluctuations of a MSN
at room temperature for increasing value of the external current.
Ohmic regime: curve (1) $I=5$ mA; non-linear-regime: curve (2) $I=200$ mA,
curve (3) $I=220$ mA and curve (4) $I=250$ mA. The solid grey lines are
the best-fit with power-laws, the resulting slopes in the different
regions of the spectrum are reported in the figure. For visual reasons
curve (2) has been multiplied by a factor $2 \times 10^2$, curve (3) 
by a factor  $2 \times 10^4$ and curve (4) by $1 \times 10^6$.} 
\end{figure} 


 

\begin{thebibliography}{10}
\bibitem{geo} Actually, two-dimensional resistor networks of sizes 
$N_1 \times N_2$ with $N_1 \ne N_2$, or networks with different lattice
structures can be easily considered.
\bibitem{nota} We take the correlation time associated with the fluctuations  
of the defect fraction, $\tau_p$ comparable with that corresponding to the 
fluctuations of the resistance, $\tau$, i.e. we take  $\tau_p \approx \tau$. 
Our simulations confirm this assumption.
\end{thebibliography}
\end{document}